# Programming frictionless interfaces for moiré layers

Zichong Zhang[1], Shuze Zhu*[1]

[1]Center for X-Mechanics, Department of Engineering Mechanics, Zhejiang University, Hangzhou 310000, China

*To whom correspondence should be addressed. E-mail: shuzezhu@zju.edu.cn

**Abstract**

Structural superlubricity in van der Waals layered systems holds immense promise for diverse nanoscale contacts devices and energy-efficient applications. While all-direction structural superlubricity has been widely investigated, the understanding towards the more fundamental directional structural superlubricity require further attentions. In this study, we reveal the physical origins of directional structural superlubricity, which reduces to all-direction superlubricity under certain conditions. By investigating the evolution of incomplete moiré tiles at crystalline interfaces, our general scaling approaches establish the mapping from geometry to tunable directional superlubricity, agreeing with large scale molecular dynamics simulations at both homogeneous or heterogeneous interfaces. Furthermore, diverse programmable frictionless motions of nanoflakes traveling inside double-surface nanoconfinement systems can be achieved. Our work delivers new insights into the design of ultra-low frictional interfaces for future nanoscale tribology and nanoconfinement transport.

Ultralow friction at incommensurate crystalline interfaces offers significant potential for reducing energy dissipation [1-6]. Such phenomenon, known as structural superlubricity, has been extensively studied in van der Waals (vdW) layered materials, with particular attention being paid to the close relationship between moiré patterns evolution and frictional properties [7-9]. The basic understanding is that the change in the collection of interface moiré tiles affects the sliding energy barrier [10-12]. The complete moiré tiles in particular cases lead to ultra-low friction in all directions [13-19]. However, the concept of all-direction superlubricity is essentially a subcategory of the more general concept of directional structural superlubricity, in which static friction vanishes in one specific direction and remains finite in other directions. Nevertheless, the mechanistic understandings and designing principles of the more fundamental directional superlubricity are not fully explored yet. Furthermore, most available superlubricity investigations are focused on the frictional characteristics over a single surface. However, in practical scenarios, the motion of nanostructure within double-surface nanoconfinement channels is frequently encountered [20-22]. For example, graphite bilayer channels with precisely-controlled Angstrom (Å) scale channel heights (e.g., 6.8 Å) have attracted considerable attention [23], and can potentially be used for low-friction molecular mass transportation [24]. Nevertheless, the designing principles of frictionless contact in double-surface nanoconfinement channel is unexplored.

In this paper, we unveil the quantitative mapping from the evolution of moiré tiles to directional structural superlubricity, which reduces to all-direction superlubricity in certain circumstances. Our scaling theory also enables the programming of superlubricity in double-surface nanoconfinement channels, which has garnered experimental evidences. Particularly, the non-superlubricity, directional superlubricity, and all-direction superlubricity are programmable within nanoconfinement channels. The theoretical understandings strongly agree with those from molecular dynamics simulations at both homogeneous and heterogeneous interfaces. Our findings not only provide rational design of nanoscale ultra-low frictional interfaces, but also open new ways into the theoretical guideline for nanoconfinement transport and nanomaterial synthesis.

We first perform the molecular dynamics [25] simulations on directional superlubricity at two representative interfaces. Fig. 1(a)-1(c) correspond to twisted graphene flake on a bottom graphene layer, which represents friction at homogeneous interface. The bottom layer is fixed and considered infinitely large. The top graphene flake, carrying the shape of a parallelogram with an acute angle of 60°, is tailored to form an incomplete moiré tile. The base side of the parallelogram, with a length of moiré periodicity $\lambda$ (95 Å) due to a relative twist angle of $\theta = 1.47°$, passes through the centers of two AA stacking domains (red), so that it coincides with one of the moiré lattice vectors $\overrightarrow{R_M}$. The height of the parallelogram is $\frac{\sqrt{3}\lambda\xi}{2}$, where $\xi$ is a size parameter. In Fig. 1(a) $\xi = \frac{2}{3}$ and the flake shape is precisely two-thirds the size of a complete moiré tile. The tailored graphene flake is dragged [26-28] to translate in different directions, while the maximum friction $F_{max}$ is recorded [14]. The results [Fig. 1(b)] show that $F_{max}$ smoothly varies as translational direction evolves. The maximum friction direction is perpendicular to the minimum friction direction. For convenience, we assign the direction of minimum friction as 0° ($x$-axis). Later, our theory will reveal that it is in fact a frictionless direction. From the moiré evolution of the flake during minimum friction sliding [Fig. 1(c)], the moiré pattern translates exactly along the base side of parallelogram with length $\lambda$, which aligns with moiré lattice vector $\overrightarrow{R_M}$. Such moiré translation direction is a consequence when the flake slides along the direction of the bottom layer lattice vector $\overrightarrow{R_b}$ [29], which we define as the frictionless direction ($x$-axis).

The directional superlubricity for graphene on hBN, which represents friction at heterogeneous interface, is shown in Fig. 1(d)-1(f). The relative twist angle for graphene flake on hBN is 0.6°. The graphene flake is also cut to a parallelogram with the base side length of moiré periodicity $\lambda$ (118 Å) [Fig. 1(d)]. However, here the geometric requirement is that such base side must pass through the centers of two AB stacking domains, which have the lowest stacking energy [30]. The height of the parallelogram flake is still denoted $\frac{\sqrt{3}\lambda\xi}{2}$ and $\xi = \frac{2}{3}$ in Fig. 1(d). Similar direction-dependent friction behavior for graphene/hBN heterostructures from simulations is illustrated [Fig. 1(e)]. Again, the sliding direction of minimum friction [Fig. 1(f)] is along the bottom layer lattice vector $\overrightarrow{R_b}$, so that the corresponding moiré pattern translates along the parallelogram base side ($\overrightarrow{R_M}$).

Our theoretical understanding to rationalize the above simulated directional superlubricity originated from moiré evolution. Fourier series are adopted to describe moiré energy [31]. We define $p$ as a ratio parameter that quantifies lattice mismatch (i.e., $p = |\vec{R_t}|/|\vec{R_b}|$), where $\vec{R_t}$ stands for lattice constant of the top sliding layer while $\vec{R_b}$ for the bottom layer. Both homogeneous interface ($p = 1$) and heterogeneous interface ($p \neq 1$) are considered in our theory. The initial moiré pattern $\rho_{init}(x', y', \lambda)$ is an energy density function in the form of plane wave superposition [Eq. (1)], where $\eta = \pm 1$ represents the nature of moiré interlayer energy distribution. For example, $\eta = +1$ for generic twisted bilayer graphene moiré system, and $\eta = -1$ for graphene/hBN moiré structure [14].

$$\rho_{init}(x', y', \lambda) = \eta \sum_{\vec{G}} \cos(\vec{G} \cdot \vec{r}) = \eta \left[\cos\left(\frac{4\pi x'}{\sqrt{3}\lambda}\right) + 2\cos\left(\frac{2\pi x'}{\sqrt{3}\lambda}\right)\cos\left(\frac{2\pi y'}{\lambda}\right)\right]. \quad (1)$$

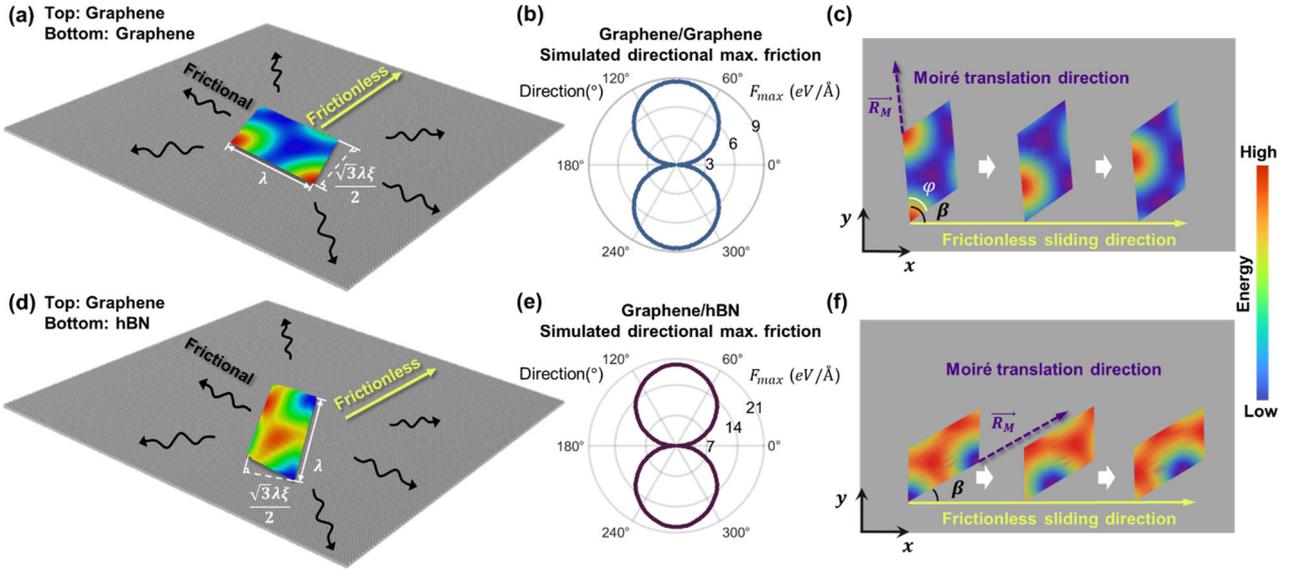

FIG. 1. Evidences and fundamental moiré physics of directional superlubricity. (a) Schematic for graphene/graphene. The top parallelogram sliding flake has a base side with the length of moiré periodicity ($\lambda$). The height from base side is $\frac{\sqrt{3}\lambda\xi}{2}$. Here $\xi = \frac{2}{3}$ is used as an example representing incomplete moiré tile. Yellow and black arrows indicate frictionless and frictional sliding direction respectively. The flake is colored by interface energy, with the red regions denote AA stacking domains. (b) Polar plot of maximum friction $F_{max}$ versus sliding direction from molecular dynamics simulations. (c) Moiré evolution within graphene flake. The frictionless sliding (yellow arrow) on bottom layer corresponds to moiré translating ($\vec{R_M}$) along base side. (d-f) Illustration for graphene/hBN. Here the blue region denotes energy favorable AB stacking domain. (c) and (f) are plotted in bottom layer coordinate system, with $x$-axis aligned with bottom layer lattice vector $\vec{R_b}$.

The energy density function in Eq. (1) uses a moiré coordinate system, where $y'$-axis is aligned with moiré superlattice vector $\vec{R_M}$, which is also parallel with the base side of the parallelogram flake. However, in our theory all vector components and coordinates $(x, y)$ are defined in the bottom layer coordinate system ($\vec{R_b}$ as $x$-axis and denoted in Fig. 1(c)). Therefore, it is essential to transform $\rho_{init}(x', y', \lambda)$ into $\rho(x, y, \theta, p)$ by a rotation matrix $R(\beta)$, where the angle [29] $\beta$ between $\vec{R_b}$ and $\vec{R_M}$ depends on $\theta$ and $p$. The moiré periodicity $\lambda$ could be given as [29] $\lambda(\theta, p=1) = \frac{|\vec{R_b}|}{2\sin\frac{\theta}{2}}$ and $\lambda(\theta, p \neq 1) = \frac{p|\vec{R_b}|}{\sqrt{1+p^2-2p\cos\theta}}$. Furthermore, the crucial physics reside in the evolution of $\rho(x, y, \theta, p)$ as the flake translation occurs. If the translation vector of the flake $\vec{v} = (\Delta x, \Delta y)$ decomposes into $x_1(\Delta x, \Delta y)\vec{R_{b1}} + x_2(\Delta x, \Delta y)\vec{R_{b2}}$, then moiré pattern translates as [29,32] $\vec{V} = x_1(\Delta x, \Delta y)\vec{R_{M1}}(\theta, p) + x_2(\Delta x, \Delta y)\vec{R_{M2}}(\theta, p)$. Finally, the translated energy density function is a function of $\Delta x$ and $\Delta y$, denoted as $\rho(x, y, \theta, p, \Delta x, \Delta y)$.

For any flake shape, we can obtain the scaling law of translational interface energy by integrating $\rho(x, y, \theta, p, \Delta x, \Delta y)$ over the initial flake region $\Omega$. Such integrating method has been introduced in our previous works [31,33]. Consider a general parallelogram shape [Fig. (1)], with base side $\lambda$, height $\frac{\sqrt{3}\lambda\xi}{2}$ and the angle between the two sides $\varphi$. The translational interface energy of the parallelogram could be analytically acquired from $\Delta E(\xi, \lambda, \Delta x, \Delta y) = \iint_{\Omega(x,y,\lambda)} \rho(x, y, \theta, p, \Delta x, \Delta y) dA$ as

$$\Delta E = \frac{\sqrt{3}\eta\lambda^2}{2\pi}\cos\left(\xi\pi + \frac{4\pi\Delta y}{\sqrt{3}\,|\vec{R_b}|}\right)\sin\xi\pi. \qquad (2)$$

Eq. (2) shows that the translational interface energy is proportional to $\sin\xi\pi$, quadratic in $\lambda$, fluctuates as $\cos\left(\xi\pi + \frac{4\pi\Delta y}{\sqrt{3}|\vec{R_b}|}\right)$ with period equaling to $\frac{\sqrt{3}}{2}|\vec{R_b}|$, but irrelevant with $\varphi$.

The frictional force for any sliding direction $l = <\cos\gamma, \sin\gamma>$, where $\gamma$ measures from the $x$-axis in bottom layer coordinate system, could be derived by directional derivative $f = -\partial\Delta E/\partial l$ as

$$f = \frac{2\eta\lambda^2 \sin\gamma \sin\xi\pi \sin\left(\xi\pi + \frac{4\pi\Delta y}{\sqrt{3}\,|\vec{R_b}|}\right)}{|\vec{R_b}|}. \qquad (3)$$

The maximal frictional force $F_{max} = \max(f)$ along sliding direction $l = <\cos\gamma, \sin\gamma>$ is

$$F_{max} = \frac{2\lambda^2}{|\vec{R_b}|}\sin\xi\pi \sin\gamma \qquad (4)$$

Eq. (3) rigorously prove that superlubricity can be achieved via either $\sin\xi\pi = 0$ (i.e., all-direction superlubricity), or $\sin\left(\xi\pi + \frac{4\pi\Delta y}{\sqrt{3}\,|\vec{R_b}|}\right) = 0$ (i.e., directional superlubricity). The maximal friction could be calculated in different direction from Eq. (4), and reaching maximum at $\gamma = \frac{\pi}{2}$ measured from $x$-axis.

Eqs. (2-4) are our main theoretical results.

When $\xi$ is an integer (i.e., complete moiré tiles), frictional force $f = 0$ along any translation direction, leading to all-direction frictionless behavior. Such strict proof on all-direction superlubricity is missing in previous studies [7,8,10,13-15,34,35]. In fact, since $\varphi$ does not appear in Eq. (3), any parallelogram of arbitrary $\varphi$ with the height of $\frac{\sqrt{3}\lambda\xi}{2}$ and base side $\lambda$ will exhibit all-direction frictionless behavior. Therefore, all-direction frictionless behavior can exist even for non-rhombic flake shape.

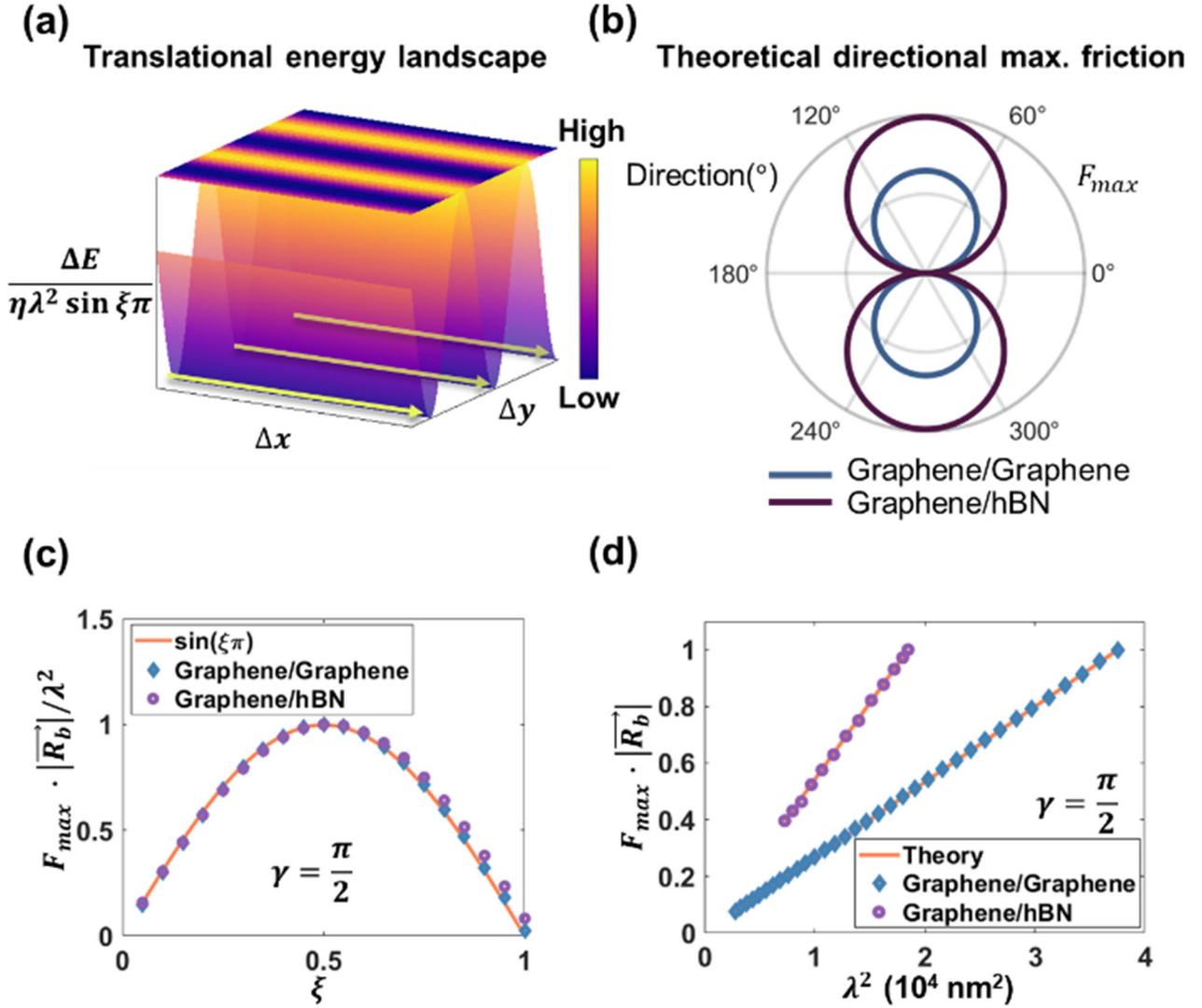

FIG. 2. Scaling laws on translational energy landscape and maximum friction force $F_{max}$. (a) Continuous zero-gradient valley paths (along yellow arrows) exist in translational energy landscape, corresponding to directional superlubricity. (b) Polar plot of theoretical scaling on $F_{max}$ versus sliding direction for graphene/graphene and graphene/hBN systems. (c) Comparisons of simulated and theoretical $F_{max} \cdot |\vec{R_b}|/\lambda^2$ with different size parameter $\xi$ for graphene/graphene and graphene/hBN. (d) Comparisons of simulated and theoretical $F_{max} \cdot |\vec{R_b}|$ with $\lambda^2$ (the square of moiré periodicity) for graphene/graphene and graphene/hBN. For (c) and (d), $F_{max}$ for $\gamma = \frac{\pi}{2}$ is plotted.

From Eq. (3), directional superlubricity requires $\sin\left(\xi\pi + \frac{4\pi\Delta y}{\sqrt{3}|\vec{R_b}|}\right) = 0$ given $\xi$ is not an integer (i.e., parallelogram flake shape hosting incomplete moiré tile), leading to $\Delta y = \frac{\sqrt{3}}{4}|\vec{R_b}|(k - \xi)$ where $k$ is an integer. For given $k$, constant $\Delta y$ is achieved by just moving along $x$ direction, which is the bottom layer $\vec{R_b}$ direction. Moreover, $\eta \cos(k\pi) = -1$ is required to ensure that directional superlubricity is at local energy minima, as illustrated by the yellow arrows in Fig. 2(a), representing continuous zero-gradient valley path. In other words, we define directional superlubricity as energetically favorable only along superlubricity direction, with energy barriers in other directions. The maximal frictional force (Eq. (4)) is sinusoidal as a function of translation direction, which in polar coordinates [Fig. 2(b)] resembles the simulation results [Fig. 1(b),1(e)]. To further validate the accuracy of Eq. (4), we quantitatively reproduce the simulation results as a function of $\xi$ [Fig. 2(c)] and $\lambda$ [Fig. 2(d)] for $\gamma = \pi/2$. $F_{max}$ reaches its largest value with $\xi = 0.5$ [Fig. 2(c)], corresponding to parallelogram flake shape hosting half of complete moiré tile. More importantly, $F_{max}$ scales linearly with the $\lambda^2$, in excellent agreement with simulation results [Fig. 2(d)]. Finally, the frictionless direction can be obtained by clockwise rotation $\beta$ from base side [as shown in Fig. 1(c), 1(f)], whose formula is

$$\beta = \begin{cases} \frac{\pi + \theta}{2} & (p = 1) \\ \arccos\left(\frac{\cos\theta - p}{\sqrt{1 + p^2 - 2p\cos\theta}}\right) & (p \neq 1) \end{cases} \quad (5)$$

More importantly, the above findings can be extended to achieve superlubricity in large-size flake, helpful for design of macroscale superlubricity. For any $\xi > 1$, the flake region can be demarcated by regions possessing complete (integer $\xi_0$) and incomplete (non-integer $\xi_1$) moiré tiles. The complete moiré tiles do not contribute to frictional force ($F_{max}(\xi_0) = 0$), so that the directional superlubricity of the entire flake $F_{max}(\xi) = 0$ requires $F_{max}(\xi_1) = 0$, which could be realized by assembling incomplete moiré tiles along the moiré superlattice vector $\vec{R_M}$. Therefore, the complete and incomplete moiré tiles can be programmed as "Lego bricks" to construct large-scale superlubricated flake. Furthermore, superlubricity in large-size flake can benefit from the enhanced rotational stability possessing multiple complete and incomplete moiré tiles.

The continuous zero-gradient valley paths in the translational energy landscape is the essential requirement for any kind of superlubricity. For example, following the same theoretical treatment, we find that flakes of equilateral triangles with side lengths equal to $\lambda$ do not possess any kind of superlubricity, because its energy landscape consists of arrays of peaks and valleys. Although there are straight line paths with $\partial \Delta E/\partial l = 0$, they are not located at zero-gradient valleys.

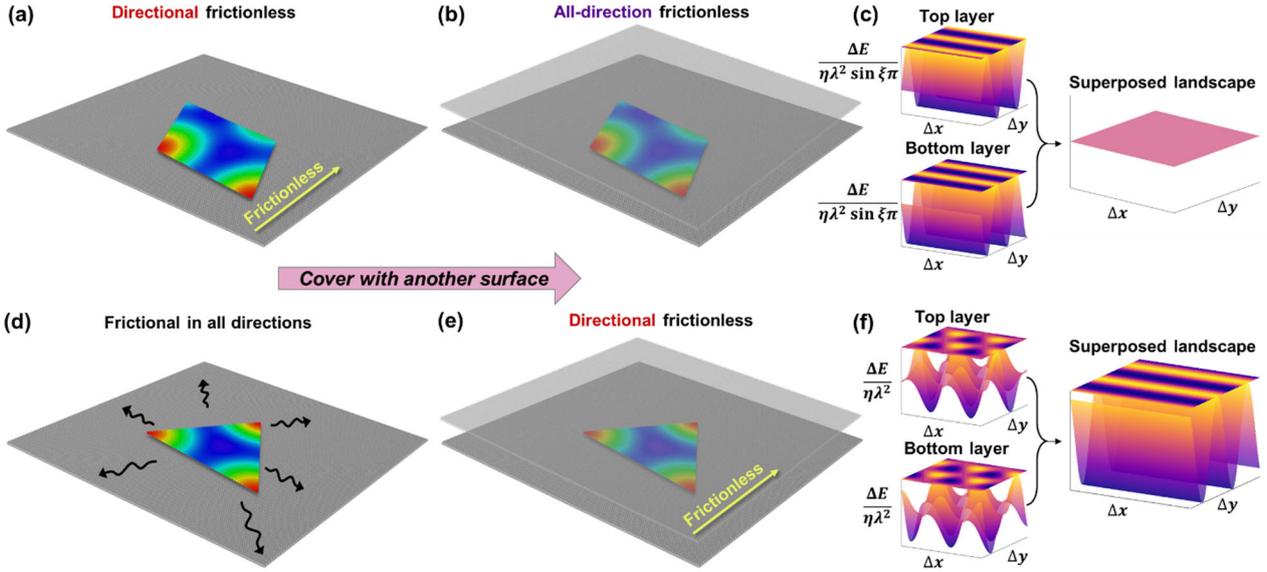

FIG. 3. Rational design of superlubricity in double-surface nanoconfinement channels. (a-c) Schematics of converting directional superlubricity (a) into all-direction superlubricity (b) for parallelogram flake, which originates from the complete flat translational energy landscape (c) from the superposed contribution with top and bottom layer. (d-f) Schematics of converting all-direction frictional behavior (d) into directional superlubricity (e) for triangular flake, which originates from the continuous zero-gradient valley paths in translational energy landscape (f) from the superposed contribution with top and bottom layer.

Our theory allows the programming of much richer superlubricity properties in bilayer nanoconfinement channels, based on superlubricity over a single surface. Particularly [Fig. 3], directional superlubricity on single surface can be promoted to all-direction superlubricity in bilayer channel, and non-superlubricity on single surface can be promoted to directional-superlubricity in bilayer channel. Depending on the shape of the sliding flake, the in-plane relative positions of the channel bottom and channel top layers can be tuned to program desirable friction behaviors. We denote the relative displacement of the channel top layer with respect to the channel bottom layer as the vector $\vec{d} = (x_0, y_0)$ in units of $|\overrightarrow{R_{bot}}|$. Still, we present our results in channel bottom layer coordinate system, which means $x$-axis is aligned with channel bottom layer lattice vector $\overrightarrow{R_{bot}}$. AA stacking corresponds to $\vec{d} = (0,0)$ and AB stacking corresponds to $\vec{d} =$

$(0, -\frac{1}{\sqrt{3}})$ for channel layers.

Following preceding discussions, parallelogram of graphene flake possessing incomplete moiré tile exhibits directional superlubricity [Fig. 3(a)]. Next, an additional top graphene layer is placed [Fig. 3(b)] so that the flake travels in the nanoconfinement channels, with $\vec{d} = \left(0, \frac{\sqrt{3}}{4}\right)$. Such vector $\vec{d}$ affects the friction property of the flake by ensuring the superposed translational interface energy to be completely flat [Fig. 3(c)], giving rise to all-direction superlubricity. Furthermore, it is also possible to promote non-superlubricity (frictional in all directions) behavior into directional superlubricity, using equilateral triangular flake with side lengths equal to $\lambda$ as examples. On a single surface system, there is no superlubricity direction [Fig. 3(d)]. However, directional superlubricity appears in nanoconfinement channels [Fig. 3(e-f)] given suitable vector $\vec{d}$ [Table 1].

TABLE I. Vectors $\vec{d} = (x_0, y_0)$ for rational design of superlubricity in nanoconfinement channels.

| Flake shape | Single surface property | Double surfaces property | Design of $(x_0, y_0)$ in units of $|\overrightarrow{R_{bot}}|$ |
|---|---|---|---|
| parallelogram | Directional superlubricity (along $0°$) | All-direction superlubricity | $\left(0, \frac{\sqrt{3}}{4}\right)$ |
| equilateral triangle (side length $\lambda$) | Non-superlubricity | Directional superlubricity (along $0°$) | $\left(0, \frac{\sqrt{3}}{2}\right)$ or $\left(\frac{1}{2}, 0\right)$ |
| equilateral triangle (side length $\lambda$) | Non-superlubricity | Directional superlubricity (along $60°$) | $\left(\frac{1}{4}, \frac{\sqrt{3}}{4}\right)$ |
| equilateral triangle (side length $\lambda$) | Non-superlubricity | Directional superlubricity (along $-60°$) | $\left(\frac{1}{4}, \frac{3\sqrt{3}}{4}\right)$ |

In conclusion, moiré-based geometric theory is developed to program frictionless interfaces. The origins of all-direction superlubricity and directional superlubricity are unified with our theoretical formulations. The continuous zero-gradient interface energy valley path is the defining feature for directional superlubricity, which reduces to all-direction superlubricity for complete moiré tile. Directional superlubricity always happens along the bottom layer lattice vector direction, with geometric requirement on the shape of the flake. Furthermore, much richer superlubricity properties in bilayer nanoconfinement channels can be programmed from the superlubricity properties over a single surface. Our findings not only provide quantitative design of ultra-low frictional interface, but also reveal possibilities to utilize superlubricity in nanoconfinement directional transport.